\documentstyle[twocolumn,floats,aps,epsf,times]{revtex}

\begin{document}

\newcommand{\be}{\begin{eqnarray}}
\newcommand{\en}{\end{eqnarray}}
\newcommand{\mc}{\mathcal}
\newcommand{\no}{\nonumber}
\newcommand{\ie}{\mbox{\protect{\it i.\ e.\ }}}
\newcommand{\eg}{\mbox{\protect{\it e.\ g.\ }}}
\newcommand{\cf}{\mbox{\protect{\it c.\ f.\ }}}
\newcommand{\etal}{\mbox\protect{\it et.\ al.\ }}

\twocolumn[
\hsize\textwidth\columnwidth\hsize
\csname@twocolumnfalse\endcsname

\title{
Orthogonality Catastrophe Revisited
}
\author{Nic Shannon}
\address{
Department of Physics, University of Wisconsin--Madison, 
1150 Univ. Av., Madison WI 53706--1390, USA.
}
\date{\today}
	
\maketitle

\begin{abstract}
We prove a simple theorem on the overlap of the wavefunctions
of a manybody system with and without a single impurity 
and show how, and under which conditions, this leads to the 
``Orthogonality Catastrophe'' (OC) described by Anderson.
A compact new derivation of the known result for the OC exponent 
in a free electron gas, 
$\alpha=2 \sum_l(2l+1) (\delta_l(\epsilon_f)/\pi)^2 $
is given. We also introduce a simple way of calculating core 
level photoemission (XPS) lineshapes from the finite size scaling 
of groundstate energies, reproducing and extending results 
previously derived using boundary conformal field theory.
Different conditions under which the OC fails are 
discussed, and simple physical predictions for XPS lineshapes made 
in each case.
\end{abstract}

\pacs{PACS ??.??}
]
\narrowtext

The problem of a single localized electronic state
--- prototypically a core orbital --- embedded in a many body system (of electrons or 
phonons) occurs in many different contexts.   
The simplest case, in which the state
couples to other excitations through Coulomb interaction alone, can be 
recast as an impurity problem, and transitions involving 
the localized level occur at rates determined by matrix elements 
between states of the manybody system with and without the impurity.  
The spectral properties of such localized states can be measured 
by X--Ray photoemission (XPS) and provide a powerful, if indirect, 
probe of the (low energy) charge carrying excitations of the manybody 
system in which it resides.  

An example of how this manybody effect comes into play is
provided by XPS spectra of core levels in many simple metals. 
The expected sharp symmetric peak at the binding energy $\epsilon_d$ 
of the core level is converted into a powerlaw singularity of the form 
$
P(\omega) \sim 1/(\omega - \epsilon_d^{\prime})^{1-\alpha}
$, 
where $\alpha \sim 0.1$, broadened by the finite lifetime of the core 
hole to give an ``asymmetric Lorentzian'' peak \cite{sunjic}).   
This powerlaw behavior was predicted by Anderson 
\cite{anderson} on the basis of the {\it Orthogonality Catastrophe} 
(OC) --- the observation that
the overlap of the groundstates of a free electron gas with and
without a single impurity vanish as 
$|O_0| \sim 1/L^{\alpha/2}$
in the thermodynamic limit $L \to \infty$, where $L$ is the 
largest characteristic length scale of the system.

The OC has proved an extremely robust and important piece of physics, 
with consequences for metallic systems reaching far beyond the simple 
``single impurity'' problem discussed above.  It and related issues 
have been treated exhaustively by many authors --- for a review see \eg \cite{ogata}.
In this paper we address the question of what new features might be 
found in the spectra of core levels coupled to many electrons systems 
unlike conventional metals, by looking considering
several scenarios in which the OC can fail.
Each of these has different and potentially observable consequences
for XPS spectra.  

We will work with the simple model Hamiltonian
\be
\label{eqn:model}
{\mc H}= {\mc H}_0 
  + d^{\dagger}d \epsilon_d
  + dd^{\dagger}{\mc V}
\en 
where $d^{\dagger}$ is the creation operator for a localized
state of binding energy $\epsilon_d$ and ${\mc V}$ is an impurity 
potential of the form
\be
\label{eqn:V}
{\mc V} = \int dx V(x) \rho(x)
\en
($\rho$ is the density of electrons).
XPS will be treated in the sudden approximation,
within which the photoelectron yield $P(\omega)$ is determined by 
the spectral function of the localized state
$
A_h(\omega) = \sum_n \mid \langle n \mid d \mid 0 \rangle \mid^2 
  \delta(\omega - \epsilon_n + \epsilon_0)
$
where $\mid 0 \rangle$ is the groundstate of the entire system
with the localized state filled, and $\{ \mid n \rangle \}$
a complete set of eigenstates for the system with the localized state empty.

We begin by giving a proof of oft cited
fact that the overlap between the ground state of the 
many electron system without the impurity and the eigenstates of 
the system with the impurity present are exponentially small 
in the number of excitations which separate them.
This leads directly to an understanding of both the necessary 
conditions for the OC to occur and a new way of calculating 
XPS lineshapes representative of the cases in which it fails.
Our treatment provides a bridge between the physical picture of OC
effects put forward by Hopfield \cite{hopfield} 
and a number of powerful results previously found by the use 
of boundary conformal field theory (BCFT) \cite{affleck}.

{\it A theorem on the overlap of manybody wavefunctions.}
We consider a many electron system initially in its
groundstate ${\mc H}_0 \mid \Psi_0 \rangle = E_0 \mid \Psi_0 \rangle$.
At time $t=0$ an impurity potential ${\mc V}$ is suddenly ``switched on''.
The suddenness of the switching implies that the system remains in the 
state $\mid \Psi_0 \rangle$, which in general is no longer an 
eigenstate and evolves instead according to ${\mc H} = {\mc H}_0 + {\mc V}$.  

We may express this sudden switching as :
\be
\langle \Psi(t=0^{-}) \mid \Psi(t=0^{+}) \rangle = 
\langle \Psi_0 \mid \Psi_0 \rangle=1
\en
Inserting a complete set of states 
${\mc H} \mid m \rangle = \epsilon_m \mid m \rangle$
(at this stage for clarity we will drop all quantum numbers except
the principle quantum number $m$ --- they may be restored at will)
and writing $O_m = \langle m \mid \Psi_0 \rangle$ we find
\be
\label{eqn:complete}
\mid O_0 \mid^2 + \sum_{m>0} \mid O_m \mid^2 = 1 
\en
where we have singled out the overlap $O_0$ of the old and 
new ground states.

We now introduce two (real, positive) functions, 
\be
\mid O(\epsilon_m) \mid = \mid O_m \mid 
   \qquad 
D(\epsilon) = \sum_m \delta (\epsilon_m - \epsilon)
\en
which respectively characterize the overlap of the ground state 
with the $m^{th}$ eigenstate of the perturbed system, and 
the density of excitations created by the impurity.
Both functions are assumed to be at least piecewise continuous and smooth.  
We then recast Eqn. (\ref{eqn:complete}) as
\be
\mid O_0 \mid^2 = 1 - \int_{\epsilon_1}^{\infty} 
  d\epsilon D(\epsilon) \mid O(\epsilon) \mid^2
\en
where $\epsilon_1$ is the lowest excitation 
energy of the system.  In order to be able to make contact 
with field theory we will assume that the smallest energy spacing
is inversely proportional to the length of the system --- 
$\epsilon_1 = (2)\pi v/L$, where $v$ is a characteristic velocity 
--- throughout this article \cite{bc}.
The ground state of the perturbed system is taken to be unique, and to
have the same overlap with the ground state as the unperturbed
system as the first excited state ($O_0 = O_1$).  This follows 
from the smoothness of $O(\epsilon)$ if we restrict ourselves to the 
limit $\epsilon_1 \to 0$.   We will consider the case of finite $\epsilon_1$ 
separately bellow.  

We can then differentiate formally on $\epsilon_1$ to obtain
\be 
\frac{\partial \mid O(\epsilon)\mid^2}{\partial \epsilon}
   = D(\epsilon) \mid O(\epsilon)\mid^2
\en
and impose the physical ``boundary condition'' 
$\mid O(\infty) \mid = 1$
to obtain the solution   
\be
\label{eqn:result1}
\mid O(\epsilon)\mid^2 = e^{-N(\epsilon)} 
   \qquad
N(\epsilon) =   \int_{\epsilon}^{\infty} d\epsilon D(\epsilon)
\en
where $N(\epsilon) \geq 0$ is a monotonically decreasing real function
subject to the boundary condition $N(\infty) = 0$, and has
the interpretation of the number of excitations excited by the 
sudden switching of the impurity with energy greater than
$\epsilon$.   The overlap between the two ground states
is then exponentially small in the total number of excitations 
produced by the impurity.  This is the theorem.

It is very easy at this stage to restore the quantum numbers for 
conserved quantities in the problem (for the free electron gas, 
different angular momentum scattering channels).   The overlap
factorizes and Eqn. (\ref{eqn:result1}) then applies separately 
in each channel, \ie 
\be
\mid O(\epsilon)\mid^2 = e^{-\sum_l n_l N_l(\epsilon)}   
\en
where $n_l$ is the  degeneracy associated with
the $l^{th}$ channel  

{\it The Orthogonality Catastrophe.}
It is easy now to show that the Anderson OC is a special 
case of the result Eqn. (\ref{eqn:result1}); however we must first
find a way of characterizing $D(\epsilon)$.   We first consider the case 
where this is a analytic function, with leading behaviour $1/\epsilon$ in 
the limit $\epsilon \to 0$ :
\be
\label{eqn:important}
D(\epsilon) = \frac{\alpha}{\epsilon} + \beta + \gamma \epsilon + \ldots
\en
This form can be shown to be correct for the free electron gas for any
sensible choice of $V(x)$; consequences 
of deviations from this behaviour will be considered separately below.

It follows immediately from the definition
Eqn. (\ref{eqn:result1}) that  
\be
N(\epsilon_1) \sim -\alpha \ln 
  \left(\frac{\epsilon_1}{\overline{\epsilon}}\right)
  + \Delta N (\epsilon_1)
\en 
where $\overline{\epsilon} = 2\pi v/a$ is an ultraviolet cutoff of order 
the bandwidth.  If in the limit $\epsilon_1 \to 0$ ($L \to \infty$),
$\Delta N (\epsilon_1)  \to \Delta N$, a constant, we then recover 
the OC 
\be
\label{eqn:OC2}
\mid O_0 \mid = A\left(\frac{a}{L} \right)^{\alpha/2}
\en
where all information about the nonuniversal 
high energy physics of the problem is hidden in the cutoff scale $a$ 
and prefactor $A = e^{-\Delta N}$.   

To get from the new ground state back to the old we must
dissipate all the excitations created by the impurity.
This implies that 
\be
\label{eqn:DeltaE1}
\Delta E_0 = \int_{\epsilon_1}^{\infty} d\epsilon \epsilon D(\epsilon) 
\en
where we have neglected a contribution 
$\langle \Psi_0 \mid {\mc V} \mid \Psi_0 \rangle$ 
which is a) independent of the system size for fixed density and b) 
involves no excitations, and so plays no part in these arguments.

Substituting Eqn. (\ref{eqn:important}) in Eqn. (\ref{eqn:DeltaE1}), 
we see that in general the OC exponent is {\it defined} by
\be
\label{eqn:profound}
\alpha = -\lim_{\epsilon_1 \to 0}
   \frac{\partial\Delta E_0}{\partial \epsilon_1} 
\en
and that the existence of this limit is a sufficient condition 
for the OC to occur.
This result is valid to all orders in the impurity strength,  
and is {\it exactly} equivalent to that found by BCFT 
\cite{affleck}, usually expressed as
\be
\alpha = \frac{L\mid\Delta E_0\mid}{\pi v}
\en
where the bulk contribution to $\Delta E_0$ is assumed to have
been subtracted in advance.  

We can apply equation Eqn. 
(\ref{eqn:profound}) directly to the case of the spinfull free electron gas
by directly substituting Friedel's result for the change in groundstate
energy due to the impurity in terms of phase shifts
\be
\Delta E = 2\sum_l (2l+1) \int_0^{\epsilon_f} \frac{d\epsilon}{\pi} \delta_l(\epsilon)
\en
and keeping track of the  leading correction to the Fermi energy 
$\delta \epsilon_f = v \delta k_f = - v \delta(\epsilon_f)/L$.  
Then to leading order in $1/L$ :
\be
\alpha &=& -\lim_{\epsilon_1 \to 0}
   \frac{\partial}{\partial \epsilon_f}
   2\sum_l(2l+1)\int_0^{\epsilon_f} \frac{d\epsilon}{\pi}
   \delta_l(\epsilon)\times
   \frac{\partial \epsilon_f}{\partial \epsilon_1}
   \no\\
   &=& 2 \sum_l(2l+1) \left[ \frac{\delta_l(\epsilon_f)}{\pi}\right]^2
\en
as found by Nozi\'eres and de Dominicis \cite{nozieres}.

In this and many other cases we can obtain a satisfactory {\it estimate} 
of $\Delta E_0$ from second order perturbation theory, which becomes exact 
in models for which the itinerant electrons behave collectively as simple harmonic 
oscillators.   Then,
\be
\label{eqn:DeltaE2}
\Delta E_0^{(2)} &=& \int_{\epsilon_1}^{\infty} \frac{d\epsilon}{2\pi} 
   \frac{R(\epsilon)}{\epsilon} \\
R(\epsilon) &=& -2\sum_q \mid V_q \mid^2 \Im \left\{ 
\chi_{\rho}^{R}(q,\epsilon) \right\}
\en
where $V_q$ is the Fourier transform of $V(x)$ and $\chi_{\rho}^{R}(q,\epsilon)$
the retarded charge susceptibility of the many electron system. 
To this order $D(\epsilon) = R(\epsilon)/2\pi\epsilon^2$. 
The function $R(\epsilon)$ {\it must} be odd and is in general analytic, \ie 
\be
\label{eqn:Repsilon}
R(\epsilon) = 
   \alpha_{\mbox{\protect{\tiny B }}} \epsilon 
   + \gamma_{\mbox{\protect{\tiny B }}} \epsilon^3 + \ldots
   \qquad \epsilon \to 0
\en
For a free electron gas perturbed by a delta function potential
of strength $V_0$, $\alpha_B = n_0^2 \mid V_0 \mid^2$, where
$n_0$ is the electron density of states at the Fermi energy, 
and the coefficient $\gamma_B$ depends on the details of the band 
structure, impurity potential, etc.   
   
{\it Failure of the Orthogonality Catastrophe.}
The OC occurs because of an infrared divergence in 
the number of excitations produced by a single impurity in 
a many electron system in the limit $L \to \infty$.
Common sense dictates that the change in ground state
energy $\Delta E_0$ due to the introduction of the impurity
must be finite.  This rules out any analytic form of $D(\epsilon)$
which diverges faster than $1/\epsilon$ in the limit $\epsilon \to 0$. 
On the other hand a logarythmic correction to Eqn. (\ref{eqn:important})
--- for example a leading term of the form 
$D(\epsilon) = \alpha \ln (\epsilon/\epsilon') / \epsilon$ ---
{\it could}  give rise to a ``hyper'' orthogonality catastrophe
in which the ground state overlap vanishes faster than $1/L^{\alpha}$,
whilst $\Delta E_0$ remained finite \cite{nicunpub}.

The orthogonality catastrophe will only fail ($O_0 \ne 0$) if the number
of excitations produced by the impurity remains countable. 
Since this means that there can be no $1/L$ term in the 
finite size corrections to $\Delta E_0$ for $L \to \infty$ it 
imposes a stringent condition on the eigenstates of the 
perturbed system; in particular there can be no simple 
electronic scattering states at the chemical potential.
Of course this second scenario cannot be precluded, and we now
consider some simple circumstances in which it applies.

We can characterize the OC in a regular metal with some generality by neglecting all 
but the leading term in Eqn. (\ref{eqn:important}) and once again imposing 
a bandwidth cutoff $\overline{\epsilon}$, which is equivalent to making the 
ansatz 
\be
\label{eqn:ansatzI}
D(\epsilon) = \frac{\alpha}{\epsilon} \theta (\overline{\epsilon}-\epsilon)
   \qquad \mbox{[Case I]}
\en
This gives a groundstate overlap $\mid O_0\mid$ scaling towards 
orthogonality according to  Eqn. (\ref{eqn:OC2}), with the 
non--universal prefactor $A=1$.

We now consider the simplest ``failure case'' for the OC, one in 
which the lowest excitation energy of the system $\epsilon_1$ remains
finite due to the existence of a gap $\Delta = 2\pi v/\xi_0$.  
We can represent this case by the ansatz :
\be
\label{eqn:ansatz2}
D(\epsilon) = \frac{\alpha}{\epsilon} 
  \theta (\epsilon - \Delta)
  \theta (\overline{\epsilon}-\epsilon)
   \qquad \mbox{[Case II]}
\en
which we denote as an ``s--wave'' gap,
and according to Eqn. (\ref{eqn:result1}) 
the overlap between the perturbed and unperturbed 
groundstates then remains finite :
\be
\mid 0_0 \mid = \left(\frac{\Delta}{\overline{\epsilon}} \right)^{\alpha/2}
   = \left(\frac{a}{\xi_0} \right)^{\alpha/2}
\en
As set out above, it is not immediately obvious that our derivation of
Eqn. (\ref{eqn:result1}) still holds for finite $\epsilon_1$, but this
can easily be understood to be the case on physical grounds.
Since there are no excitations between $\epsilon = 0$ (the 
groundstate) and $\epsilon = \epsilon_1$ there can be no change 
in the size of the overlap, \ie $\mid O_1 \mid \equiv \mid O_0 \mid$

We can also envisage a scenario in which the many electron system {\it does} have
low energy excitations, but the OC is frustrated because the number
these excited by the impurity remains countable.  
This follows trivially if the leading term in the expansion of 
$D(\epsilon\to 0)$ 
is not $\alpha/\epsilon$ but $\gamma\epsilon$, which is the case, for example, 
in a ``d--wave'' superconductor at zero temperature \cite{haslinger}.
We assume that once again there is some energy scale $\Delta$ at 
which the system crosses over to conventional ``metallic'' behavior 
and characterize this case by the ansatz
\be  
\label{eqn:ansatz3}
D(\epsilon) = \left\{ 
   \begin{array}{ll}
      \alpha \epsilon/\Delta^2
         & \quad 0< \epsilon < \Delta\\
       \alpha/\epsilon
         & \quad \Delta < \epsilon < \overline{\epsilon}\\  
      0 & \quad \epsilon > \overline{\epsilon}
   \end{array} \right.  
\en
which for simplicity we will denote as a ``d--wave'' gap.
The overlap of the groundstates is given in this case by
\be
\mid 0_0 \mid = \left(\frac{\Delta}{\overline{\epsilon}} \right)^{\alpha/2}
   e^{-\alpha/4}
   = \left(\frac{a}{\xi_0} \right)^{\alpha/2} e^{-\alpha/4}
\en
which is once again finite.  We consider the consequences
of both of these failure cases for XPS spectra below.

\begin{figure}[tb]
\begin{center}
\leavevmode
\epsfxsize \columnwidth
\epsffile{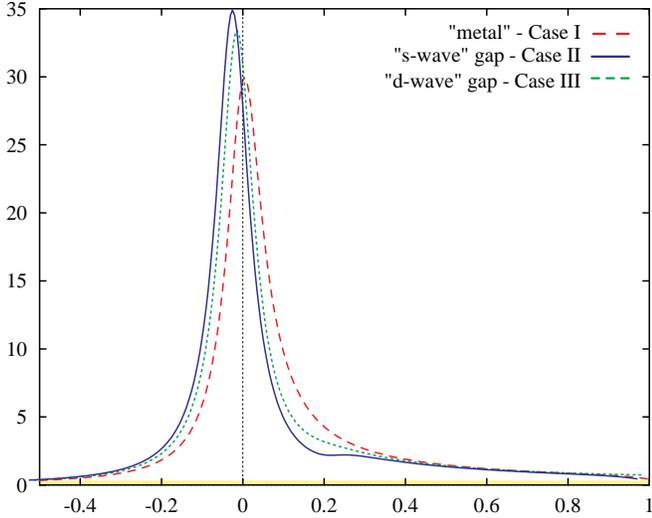}
\caption{
XPS lineshapes for a ``metal'' (Case I), 
and for the two OC failure cases
``s--wave'' gap (Case II) and 
``d--wave'' gap (Case III) made by
by convoluting the appropriate prediction for $A_h(\omega)$ 
with a Lorentzian of width $\gamma = 0.05$. 
The bare asymmetry exponent $\alpha = 0.1$ in each case, 
and the gap parameter $\Delta = 0.25$.
All energies are measured in units of the band cutoff 
$\overline{\epsilon}$.
}
\label{fig1}
\end{center}
\end{figure}

{\it Calculation of XPS lineshapes.}
The spectral function for the localized level can
be shown to depend only on the overlaps of the eigenstates
of the perturbed many electron system with its unperturbed 
groundstate and their relative energies (all of which are offset
by $\epsilon_d$).  Then, using the result Eqn. (\ref{eqn:result1}), 
we find
\be
\label{eqn:spectral}
A_h(\omega) &=& 2\pi e^{-N(\epsilon_1)} 
   \delta(\omega)\no\\
   && \quad + 2\pi \theta(\omega - \epsilon_1)
   \frac{\partial e^{-N(\omega)}}{\partial \omega}
\en
where $\epsilon_d^{\prime} = \epsilon_d + \Delta E_0$, and 
$N(\epsilon)$ is defined as above with $N(\infty) = 0$.
We will henceforth set the threshold for photoemission 
($\epsilon_d^{\prime}$) to zero.
It follows from the definition (\ref{eqn:spectral}) that the 
spectral function is {\it automatically} correctly 
normalized ($\int d\omega A_h(\omega) = 2\pi$).

The powerlaw singularity in XPS spectra associated with the OC 
can be reproduced simply by substituting the ansatz 
Eqn. (\ref{eqn:ansatzI}) into Eqn. (\ref{eqn:spectral}) to yield
\be
\label{eqn:spectrum1}
A_h (\omega)
   = \frac{2\pi\alpha}{\overline{\epsilon}} 
   \left(
      \frac{\overline{\epsilon}}{\omega}
   \right)^{1 - \alpha}
   \theta (\omega)
   \theta (\overline{\epsilon} - \omega)
\en
A mock XPS lineshape made by convoluting this prediction with 
a Lorentzian to mimic finite core hole lifetime is illustrated 
in Figure \ref{fig1}.  It is also possible to
use a soft bandwidth cutoff
($D(\epsilon) = \alpha \epsilon \exp [-\epsilon/\overline{\epsilon}]$), 
in which case we recover a normalized form of the familiar Doniach--Sunjic 
lineshape.

We now consider the consequences of the ``s--wave'' gap (Case II, 
above).  On physical grounds we anticipate that the failure of the OC 
(finite ground state overlap) leads to the 
restoration of a $\delta$--function peak at threshold.  Spectral 
weight is transferred the tail of the line to this threshold singularity 
over a range of energies corresponding to the gap.
 
The ansatz Eqn. (\ref{eqn:ansatz2}) leads directly to 
the spectral function
\be 
\label{eqn:spectrum2}
&A_h& (\omega) = 2\pi \left(\frac{\Delta}{\overline{\epsilon}}\right)^{\alpha}
    \delta (\omega) \no\\
&& + \quad
   \frac{2\pi\alpha}{\overline{\epsilon}} \theta (\omega - \Delta)
   \theta (\overline{\epsilon}- \omega)
   \left(\frac{\overline{\epsilon}}{\omega}\right)^{1 - \alpha}
\en
In general XPS lines for gapped systems are shifted to a {\it lower} binding energy
than the corresponding ``metal''; within this simple model the {\it relative} 
shift in the core line threshold due to the opening of the gap is proportional 
to the gap : 
$\Delta\epsilon_d \propto \alpha \Delta$.   
The new prediction for an XPS lineshape, including the appropriate 
shift, is illustrated in Figure \ref{fig1}.
The delta function at the new threshold energy leads to a 
marked ``sharpening'' of the line, even after finite core
lifetime has been allowed for.   Although our terminology
``s--wave'' is driven by consideration of superconducting 
systems, the parameters chosen for illustration here
may be of more relevance to systems undergoing a charge density wave 
or metal--insulator transition.

We have checked that the ansatz Eqn. (\ref{eqn:ansatz2})
reproduces the essential features of a real gapped system by calculating 
lineshapes and shifts numerically for both a semiconductor and a superconductor, 
within a more conventional perturbative second order cumulant expansion.   
The details of these 
cases will be reproduced elsewhere \cite{haslinger}, as the discussion is much
simplified by using Eqn. (\ref{eqn:spectral}) together with a simple analytic form 
for $D(\epsilon)$.
Convolution with a Lorentzian of realistic width in any case makes
corrections to all except the prefactor on the $\delta$--function 
unobservable.  

We expect the spectral function for the 
``d--wave'' gap (Case III), to behave in a broadly similar way, 
but to be shifted by a smaller amount relative to the metal and to exhibit 
additional spectral weight in the gap region $0<\omega < \Delta$, with 
correspondingly less weight in the threshold delta function.
These expectations are fulfilled by the spectral function which follows
from the ansatz Eqn. \ref{eqn:ansatz3} :
\be
\label{eqn:spectrum3}
&&A_h(\omega) = 2\pi \left(\frac{\Delta}{\overline{\epsilon}}\right)^{\alpha}
   e^{-\alpha/2} \delta(\omega)\no\\
&& + \left\{
   \begin{array}{ll} 
     \frac{2\pi \alpha \omega}{\Delta^2} 
     \left(\frac{\Delta}{\overline{\epsilon}}\right)^{\alpha}
     e^{-\alpha/2 \left[1 - \left(\frac{\omega}{\Delta}\right)^2\right]}
        &  \quad 0 < \omega < \Delta\\
     \frac{2\pi\alpha}{\overline{\epsilon}}
        \left(\frac{\omega}{\overline{\epsilon}}\right)^{1-\alpha} 
        & \quad \Delta < \omega < \overline{\epsilon}\no\\
     0  & \quad \omega > \overline{\epsilon}
   \end{array}
   \right.
\en
Once again the corresponding XPS lineshape is illustrated in Figure \ref{fig1}.

{\it Conclusions.}
Reconsidering Anderson's OC, we have shown that powerlaw scaling
towards orthogonality is only one of a number of possible outcomes 
for the groundstate of many electron system perturbed by a single impurity.
The OC may fail due to the absence of low energy excitations in the 
perturbed system or because too few of them are excited.  
By formulating an explicit relation between the scaling of the groundstate 
energy with system size and the spectral function
for the impurity state, we are able to extend the known BCFT result 
for the exponent of the core level spectral function at threshold to 
the calculation of structure in the spectral function at finite energies, 
and to explore the consequences of two simple failure cases of the OC for 
XPS lineshapes.
The predicted modification of XPS lineshapes should be measurable 
in systems undergoing charge density wave
or metal insulator transitions where gap scales are 
large compared with core hole lifetimes; the associated shift 
in lines may be measurable for superconducting systems with much 
smaller gaps.

{\it Acknowledgments.}
It is our pleasure to thank all those with whom we have discussed
these ideas; we are particularly grateful to 
Ian Affleck, Jim Allen, and Volker Meden for useful 
comments and suggestions, 
and to Robert Haslinger for help in preparing the figures. 
This work was supported under grant numbers DMR 9704972 and DMR 9632527.


\begin{thebibliography}{}
\bibitem{bc} The prefactor (2) assumes the imposition of periodic boundary conditions 
on the system.  To make contact with the boundary conditions appropriate to
scattering states (and  BCFT) it should be dropped, \ie $\epsilon_1 = \pi v/L$.
\end{thebibliography}
\end{document}